\documentclass[aps,prb,twocolumn,showpacs]{revtex4-1}
\usepackage{graphicx}
\usepackage{amsfonts,amsmath,amssymb}
\usepackage{amsthm}
\usepackage{dsfont}
\usepackage{bm}%
\usepackage{color}
\usepackage{soul} %Include when comparing comments
\usepackage{amsbsy}
\usepackage{upgreek}

\usepackage[colorlinks=true,linkcolor=blue,pagecolor=blue,filecolor=blue,menucolor=blue,urlcolor=blue,citecolor=blue,anchorcolor=blue]{hyperref}%

\newcommand{\sy}[1]{\sigma_2 }
\newcommand{\etal}{\emph{et al.}}

\newcommand{\bsigma}{{\boldsymbol \sigma}}
\newcommand{\bnabla}{{\boldsymbol \nabla}}

\newcommand{\bbeta}{{\boldsymbol \eta}}
\usepackage{amssymb,amsmath,bm}

\begin{document}

\title{Spontaneous supercurrent and $\varphi_0$ phase shift parallel
  to magnetized topological insulator interfaces}
\author{Mohammad Alidoust}
\affiliation{Department of Physics, K.N. Toosi University of Technology, Tehran 15875-4416, Iran}
\author{Hossein Hamzehpour}
\affiliation{Department of Physics, K.N. Toosi University of Technology, Tehran 15875-4416, Iran}

\date{\today}
\begin{abstract}
Employing a Keldysh-Eilenberger technique, we theoretically study the
generation of a spontaneous supercurrent and the appearance of the $\varphi_0$
phase shift parallel to
uniformly in-plane magnetized superconducting interfaces made of the surface
states of a three-dimensional topological insulator. We consider two
weakly coupled uniformly magnetized
superconducting surfaces where a macroscopic phase difference between
the $s$-wave superconductors can be controlled externally. We find that,
depending on the magnetization strength and orientation on each side,
a spontaneous supercurrent due to the $\varphi_0$-states flows
parallel to the interface at the junction location. Our calculations
demonstrate that nonsinusoidal phase relations of current components
with opposite directions result in maximal spontaneous supercurrent at phase differences close
to $\pi$. We also study the Andreev subgap channels at the interface and show
that the spin-momentum locking phenomenon in the surface states can be
uncovered through density of states studies. We finally discuss realistic
experimental implications of our findings.  
\end{abstract}


\maketitle

\section{introduction}

The topological insulator (TI) is a new state of matter with
revolutionary prospects in topological superconducting spintronics and
topological quantum computation \cite{kane_rmp,zhang_rmp}. The
topological insulators rely mainly on strong spin-orbit
couplings and possess conductive surfaces, while showing insulating aspects
in their bulk. Subsequently, TIs provide unique realistic
platforms to study robust quantum relativistic phenomena such as
spin-momentum locking and quantum spin Hall effect
\cite{kane_rmp,zhang_rmp}.

The spin direction of a moving particle at the surface of a
three-dimensional (3D) TI in the presence of time-reversal symmetry is
rigidly locked to its momentum direction \cite{kane_rmp,zhang_rmp}. 
Due to the spin-momentum locking phenomenon, the induction of superconductivity and magnetism in the surface of a TI
is predicted to serve as an unprecedented condensed matter
platform that supports odd-frequency, topological superconductivity, and Majorana
fermions \cite{Volovik,exp_ti2a, exp_ti2b,exp_ti2c2,exp_ti4a,exp_ti4b,exp_ti4c,Pientka,Vasenko,yu}. In order to fabricate
superconducting and magnetic surface states, one can
proximitize the surface channels with a superconductor and
ferromagnet, respectively
\cite{Efetov_Rev,Buzdin_Rev}. The proximity-induced superconductivity
and magnetism in the surface states are externally controllable 
through manipulating their inductors. It is worth mentioning that 
the spin-orbit coupling in the presence of
superconductivity and magnetism can result in intriguing
spin-dependent phenomena in various materials even in
systems with strong nonmagnetic scatterings \cite{buz_so1,buz_so3,berg_so2,berg_so3,bob_so1,ali_so1,ali_so2,ali_gr}. 

In recent experiments on the quantum transport through TI surface states reported by different groups, it has been concluded
that a proper theoretical framework, describing all kinds of samples
including disordered ones, is
an approach that accommodates the possibility of the presence of nonmagnetic scattering resources \cite{exp_ti4a,burkov1,exp_ti2b,Kurter_TI_exp}. To
provide such a theoretical framework, the quasiclassical approach in the equilibrium and
nonequilibrium states was generalized for 
3D TI surface states with different levels of the density of nonmagnetic impurities in
the presence of superconductivity and magnetism with arbitrary
magnetization patterns \cite{zyuzin,bobkova}. This approach was employed to study TI-based diffusive
Josephson configurations involving chiral helical magnetizations and the Edelstine phenomenon at the surface states. It was theoretically shown that
well-controlled $0$-$\pi$ supercurrent reversals, $\varphi_0$ junctions,
and proximity-induced vortices \cite{jap_alidoust} are accessible through the
spin-momentum locking phenomenon and magnetoelectric effect
\cite{zyuzin,bobkova}. Also, several works have demonstrated
that the spin-momentum locking phenomenon in the surface states of 3D TI plays crucial roles 
independently of the amount of nonmagnetic impurities present in these channels \cite{gri_1,gri_2,gri_4,houzet,fariborz,malsh,tanaka}.

The spontaneous surface flow of currents can occur in the context of
$^3$He-$A$ superfluid \cite{Volovik,volovik2}. Also,
the unconventional superconductors in proper situations host
surface states and spontaneous currents parallel to interfaces
\cite{mineev,Frusaki,bob3,ashby,sigrist1,Nandkishore,Kiesel,Tada,Bjornsson,Kirtley,volovik2,Mackenzie,Smidman}.
This phenomenon has theoretically been studied in several systems including Josephson
junctions made of $s$-wave/$d$-wave superconductors with different
crystallographic orientations, chiral superconductors, and
ferromagnetic layers coupled to the unconventional superconductors
\cite{amin,golubov,Kuboki,asano,Brydon}. Nonetheless, the
experimental observation of spontaneous supercurrents at the surfaces of chiral superconductors is still elusive
partially due to the Meissner effect and strong
disorders that may exist at the surfaces \cite{Nandkishore,Kiesel,Bjornsson,Kirtley}. 

In this paper, we make use of the Eilenberger equation derived in
Ref. \onlinecite{zyuzin} to analyze supercurrent flows at the interfaces of uniformly magnetized
superconducting surface states of 3D TIs. We consider a 2D  Josephson
weak-link configuration made of surface channels of a 3D TI with an externally controllable
superconducting phase difference $\varphi$ and uniform in-plane magnetizations
in each segment depicted in Fig. \ref{fig1}. Note that the superconductivity inductor electrodes are
spin-singlet superconductors. We show that when the magnetization strength within the left and
right sides of the contact are unequal and perpendicular to the
interface, a spontaneous supercurrent flows along the junction interface, where
its direction and amplitude can be controlled through the extrinsic
$\varphi$ applied perpendicular to the interface. Our results demonstrate
that the maximum value of the spontaneous current can be achieved at
phase differences close to $\varphi=\pi$. We justify our findings by
calculating the phase relation of
spontaneous supercurrent components along two opposite
trajectories parallel to the junction
interface. The observation of the predicted spontaneous supercurrent
is direct evidence of the rigid spin-momentum locking phenomenon in
the surface states
\cite{Kurter_TI_exp,zyuzin,bobkova,gri_1,gri_2,houzet}. Furthermore, we
calculate the density of states (DOS) at the
interface and discuss how the strength and direction of magnetizations can alter Andreev bound states in such junctions that in turn
reveal the role of strong spin-momentum locking in the surface
states of a 3D TI.

The paper is organized as follows. We first explain the setup
considered and derive proper Green's function describing the
system in Sec. \ref{results}. Next, using the Green's function obtained, we calculate the
spontaneous supercurrent along the junction interface of the Josephson
weak-link configuration and
discuss the phase relation of its components. We also calculate the DOS and the Andreev subgap
states for various values of $\varphi$ and magnetization
orientations. We support our numerical findings by Riccati-parametrizing the Green's function and deriving analytical
expressions for the Andreev bound states in different situations. 
We finally give concluding remarks in Sec. \ref{conclusions}.

\section{method and results} \label{results}

\begin{figure}[b!]
\includegraphics[width=8.50cm,height=5.70cm]{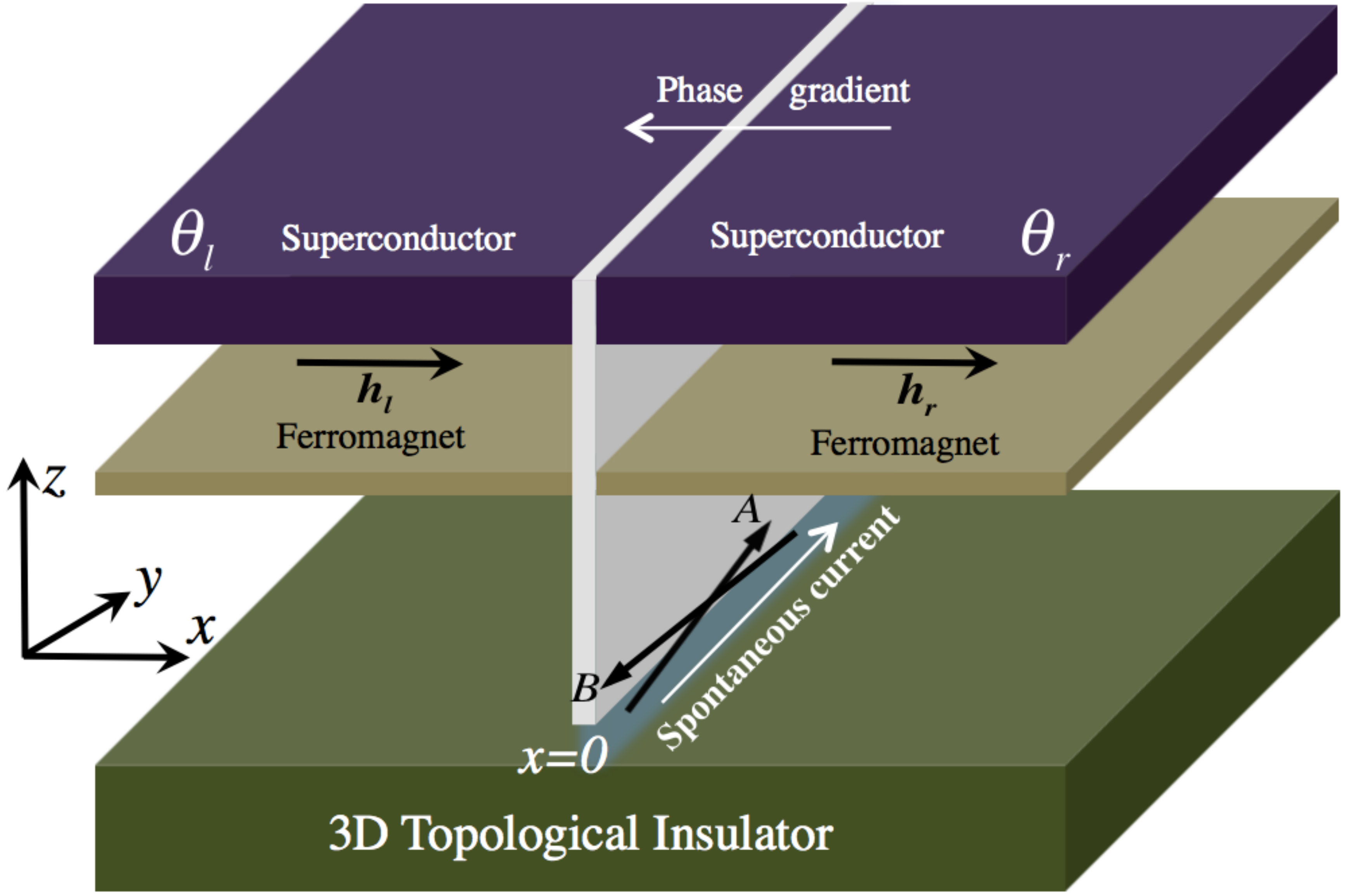}
\caption{ (Color online)  Schematic of the TI-based Josephson weak-link junction considered. The superconductivity and magnetism
  are induced in the surface states of the 3D TI by virtue of
  the proximity effect. The macroscopic phases of the left and right
  superconductors $\theta_{l,r}$ can be controlled externally. The
  orientation of the uniform in-plane magnetization induced in the surface states $\bm{h}_{l,r}$
  also can be calibrated through an external magnetic field. The
  left and right segments are separated by an insulator at $x=0$,
  constituting an interface along the $y$ axis. We assume
  that a phase gradient is applied across the junction normal to the
  interface and consider two trajectories in the $xy$ plane, parallel
  with the interface, marked by $A$ and $B$ arrows to
  analyze the supercurrent flow parallel to the junction interface
  along the $y$ axis.     
 }\label{fig1}
\end{figure}

In order to analyze the spontaneous supercurrent flow along a
uniformly in-plane magnetized interface made of surface channels of a 3D TI, we consider a Josephson weak-link
shown in Fig. \ref{fig1}. The superconductivity and magnetism are both
extrinsically induced in the surface states through $s$-wave
spin-singlet superconductors and ferromagnetic thin films, respectively, and therefore can be
calibrated externally. The Cooper pair wave-function $\Psi$ inside the ferromagnetic layer decays and oscillates as a function of location,
i.e., $\Psi\propto \Delta \exp (-z/\xi_f)\cos (z/\xi_f) $, in which
$\Delta$ is the superconducting order parameter inside the bulk
superconductor and $\xi_f$ is a characteristic length given by
$\xi_f=\sqrt{D/h}$ in a diffusive ferromagnet with the diffusive
constant $D$ and exchange field $h$ \cite{Buzdin_Rev}. Thus, the thickness of the ferromagnetic layers
should be properly chosen so that the superconductivity survives at
the surface states. The orientation of magnetization
induced in the surface channels $\bm{h}_{l,r}$ can be rotated by applying an external
magnetic field \cite{robinson_hmetal}. To fabricate the double
ferromagnetic setup depicted in Fig. \ref{fig1}, one can use different magnetic elements or
compounds that respond differently to an externally applied magnetic
field. For example, Py is a weak ferromagnet while LCMO is
strong. When subjected to an external magnetic field, the magnetization of LCMO
rotates reluctantly compared to Py \cite{robinson_hmetal} so this
constitutes favorably misaligned magnetizations. The superconducting
phase difference $\varphi=\theta_r-\theta_l$ can be controlled by passing a tuneable supercurrent through the top
superconducting electrodes ($\theta_{l,r}$ are macroscopic phases of
left and right superconductors). The two segments of the weak-link are separated
by an insulator barrier along the $y$ axis and the junction resides
at $x=0$.

To study the supercurrent flow in the weak-link Josephson structure, we follow Ref. \onlinecite{zyuzin},
where the Usadel \cite{usadel} and Eilenberger \cite{eiln}
techniques were generalized for the surface channels of a 3D TI in the presence
of superconductivity and magnetism with different amounts of
nonmagnetic impurities. Without losing the generality of our main conclusions, we utilize the Eilenberger equation \cite{zyuzin} throughout the paper:
\begin{eqnarray}\label{Eilenbrg}
&&\frac{\alpha}{2}\{\hat{\bbeta}, \bnabla \check{g}\}=\nonumber\\&&\bigg[\check{g} ,
\omega_n \hat{\tau}^z+i{\bm h}({\bm r})\cdot\hat{\bsigma}\hat{\tau}^z+i
\mu\hat{\bbeta}\cdot \mathbf{ n}_F+i\check{\Delta}({\bm r})+\frac{\langle
  \check{g} \rangle}{ \tau}\bigg],\\&&\nonumber \check{g}(\omega_n,{\bm r})=\left(\begin{array}{cc}
            g (\omega_n,{\bm r}) &f (\omega_n,{\bm r}) \\
            \tilde{f}(\omega_n,{\bm r}) & \tilde{g}(\omega_n,{\bm r})
          \end{array}\right),
\end{eqnarray}
where $\hat{\bm \bbeta} = (-\hat{\sigma}^y,\hat{\sigma}^x)$, $\hat{\tau}^{\pm} =\hat{\tau}^x\pm i
\hat{\tau}^y$, $\check{\Delta}({\bm r}) = \hat{\sigma}^0(-\Delta ({\bm r})\hat{\tau}^{+}
+\Delta^*({\bm r})\hat{\tau}^-)/2$, and ${\bm r}$ denotes the spatial
dependence of quantities. The total Green's function $\check{g}$ has four
components $f,g,\tilde{f},\tilde{g}$ that determine the physical properties of
a system. Here $\alpha$ represents the strength of the Rashba
spin-orbit coupling available at the surface channels, ${\bm h}$ is
the exchange field induced in the surface states, $\mathbf{ n}_F =
\mathbf{ p}_F /|\mathbf{ p}_F |$ is a unit vector in the direction of
momentum $\mathbf{ p}_F$ at the Fermi level,
$\omega_n=(2n+1)\pi T$ is the Matsubara frequency, $T$ is the
temperature, and $n \in Z$. The vector $\hat{\bsigma}$ is composed
of Pauli matrices and used for the spin space while $\hat{\boldsymbol
  \tau}$ denotes the particle-hole space. The parameter $\tau$ describes the
mean-free-path time of moving quasiparticles in the presence of
nonmagnetic impurities. Note that in the ballistic regime
$1/\tau\rightarrow 0$ and to simplify our calculations we neglect the term
$\langle g \rangle/\tau$ without losing the generality of our main conclusions. The
Eilenberger equation (\ref{Eilenbrg}) should be supplemented
by a normalization condition, i.e., $\check{g}\check{g}=1$, to provide
correct solutions.

To appropriately describe the physics of the interface, we consider Ansats of type
$a^i_r+b^i_re^{- k_r{\bm r}}, x>0$ and $a^i_l+b^i_le^{+ k_l{\bm r}}, x<0$ to the solutions
of the Green's function components ($i=1-4$ represents a specific component) on the right and left sides of the
weak-link, respectively \cite{ko,Kulik,Bulaevskii,golubov2}. We match the solutions
at $x=0$ where the two segments are weakly connected, derive
analytically the
corresponding Green's function, and eventually extract results
by numerically integrating over the Matsubara frequency. To obtain
$a^i_{l,r}$ and $b^i_{l,r}$ coefficients, we substitute the introduced Ansatzs into the Eilenberger
equation (\ref{Eilenbrg}), make use of the normalization condition $\check{g}\check{g}=1$,
and assume that the solutions far enough away from the interface
reduce to bulk solutions. Following this approach, we
find suitable solutions to the components of the Green's function
($x>0$):
\small
\begin{subequations}
\begin{equation}\label{g}
g_r(\omega_n)=\frac{i\omega_n+h_r}{\sqrt{(i\omega_n+h_r)^2-\Delta_r^2}}+b_re^{-k_r{\bm
    r}},
\end{equation}  
\begin{equation}\label{f}
f_r(\omega_n)=\frac{\Delta_r}{\sqrt{(i\omega_n+h_r)^2-\Delta_r^2}}+\frac{2b_r\Delta_r
e^{-k_r{\bm
    r}}}{2(i\omega_n+h_r)-i\alpha_r
k_r},~~~
\end{equation}  
\begin{equation}\label{tf}
\tilde{f}_r(\omega_n)=\frac{-\Delta_r^*}{\sqrt{(i\omega_n+h_r)^2-\Delta_r^2}}-\frac{2b_r\Delta_r^*e^{-k_r{\bm
    r}}}{2(i\omega_n+h_r)+i\alpha_r
k_r},~~~
\end{equation} 
\begin{equation}\label{tg}
\tilde{g}_r(\omega_n)=\frac{-i\omega_n-h_r}{\sqrt{(i\omega_n+h_r)^2-\Delta_r^2}}-b_re^{-k_r{\bm
    r}},
\end{equation}   
\end{subequations}
in which wavevector $ k_r= 2 \alpha_r^{-1}\sqrt{-(i\omega_n+h_r)^2+\Delta_r^2}$.
To find solutions in the left segment $x<0$, it suffices we
follow the same procedure with replacing $k\rightarrow -k$ and indices
$r\rightarrow l$. A
generic solution at the interface can be given by invoking indices $l,r$ for the
Green's function and parameters involved on the left and right sides of the junction shown in
Fig. \ref{fig1}.

By matching the Green's function of the left and right segments of the
weak-link at the junction location $x=0$, the
Green's function of the interface $\check{g}$ can be expressed
by the
following $g$ and $f$ components: 
\begin{widetext}
\small
\begin{subequations}
\begin{eqnarray}\label{normalG}
g(\omega_n)=\frac{\alpha_r k_r \Omega_l\Delta_r e^{i \theta_r} i\omega_n 
   (i\omega_n+h_l) (\alpha_l k_l-2 i (i\omega_n+h_l))+\alpha_l k_l \Omega_r \Delta_l e^{i\theta_l} i\omega_n
   (i\omega_n+h_r) (2
   (i\omega_n+h_r)-i \alpha_r k_r)}{2 \Omega_l \Omega_r (i\omega_n+h_l)
   (i\omega_n+h_r) \left\{\Delta_l e^{i \theta_l} [ 2 (i\omega_n+h_r)-i \alpha_r
   k_r ]-\Delta_r e^{i \theta_r} [ 2 (i\omega_n+h_l)+i \alpha_l
   k_l ] \right\} },
\end{eqnarray}
\begin{eqnarray}
f(\omega_n)=\frac{-\omega_n\Delta_r\Delta_l e^{i
    (\theta_r+\theta_l)}\{ \alpha_rk_r(i\omega_n+h_l)\Omega_l+\alpha_lk_l(i\omega_n+h_r) \Omega_r \}}{\Omega_l\Omega_r (i\omega_n+h_l)(i\omega_n+h_r)\{ \Delta_r e^{i\theta_r}[+i\alpha_lk_l+2(i\omega_n+h_l)]-\Delta_l e^{i\theta_l}[-i\alpha_rk_r+2(i\omega_n+h_r)]  \}},
\end{eqnarray}
\begin{eqnarray}
\Omega_{l,r}=\sqrt{\Delta_{l,r}^2+\omega_n^2},\;\;\;k_{l,r}=\frac{2}{\alpha_{l,r}}\sqrt{\Delta_{l,r}^2-(i\omega_n+h_{l,r})^2}.\nonumber
\end{eqnarray}
\end{subequations}
\end{widetext}
Because of the charge
conservation law, the Green's function at the interface, i.e., $x=0$,
is sufficient to study the supercurrent and thus we restrict our attention to
$\check{g}(x=0)$. Note that the spatial dependence of the total Green's function is given by
Eqs. (\ref{g})-(\ref{tg}).
The charge supercurrent is given through the $g$ component of
the total Green's function:
\begin{eqnarray}
{\bm J}_e({\bm r})=2ie\pi TN(0) \sum_{n\in Z}\Big<  {\bm v}_Fg(\omega_n,{\bm r})\Big>,
\end{eqnarray}
where the average $\langle ... \rangle$ is taken over the direction of momentum, ${\bm v}_F$
is the Fermi velocity, and $N(0)$ the density of states at the Fermi level.

To gain insight, let us first assume that $h_{l,r}=0$, $\theta_l=-\varphi/2$, $\theta_r=+\varphi/2$, and $\Delta_r\neq \Delta_l$. In this
case, we find the following current phase relation to the supercurrent flowing in the $x$ direction:
\begin{eqnarray}
 J_e^x=&&2e\pi TN(0) \times\nonumber\\&&\sum_{n\in Z}\frac{\Delta_r\Delta_l\sin\varphi}{\omega_n^2+\sqrt{(\omega_n^2+\Delta_r^2)(\omega_n^2+\Delta_l^2)}+\Delta_r\Delta_l\cos\varphi}.~~~~~~
\end{eqnarray}
As seen, the supercurrent in the $x$ direction is \textit{directly}
proportional to the
order parameter of the left and right superconductors, i.e., $\Delta_l$ and $\Delta_r$. Therefore, if
one of the gaps is directional dependent, for example in a
$d_{x^2-y^2}$-wave superconductor [i.e., $\Delta_{l} (\text{or } \Delta_{r})=\Delta_0\cos 2(\theta-\chi)$ where
$v_x=|{\bm v}_F|\cos\theta$ is the particle velocity in the $x$
direction and $\chi$ is the angle that the $d$-vector makes with
respect to an axis normal to the interface], we see that the
supercurrent along the $A$ and $B$ trajectories
(shown in Fig. \ref{fig1}) can be
unequal in amplitude when $\varphi\neq
0$. This implies that a finite
spontaneous supercurrent can flow along the interface when a nonzero
superconducting phase difference is applied perpendicular to the
junction interface in the $x$ direction, i.e., $\varphi\neq 0$\cite{amin}.
\begin{figure}[t!]
\includegraphics[width=7.0cm,height=6.40cm]{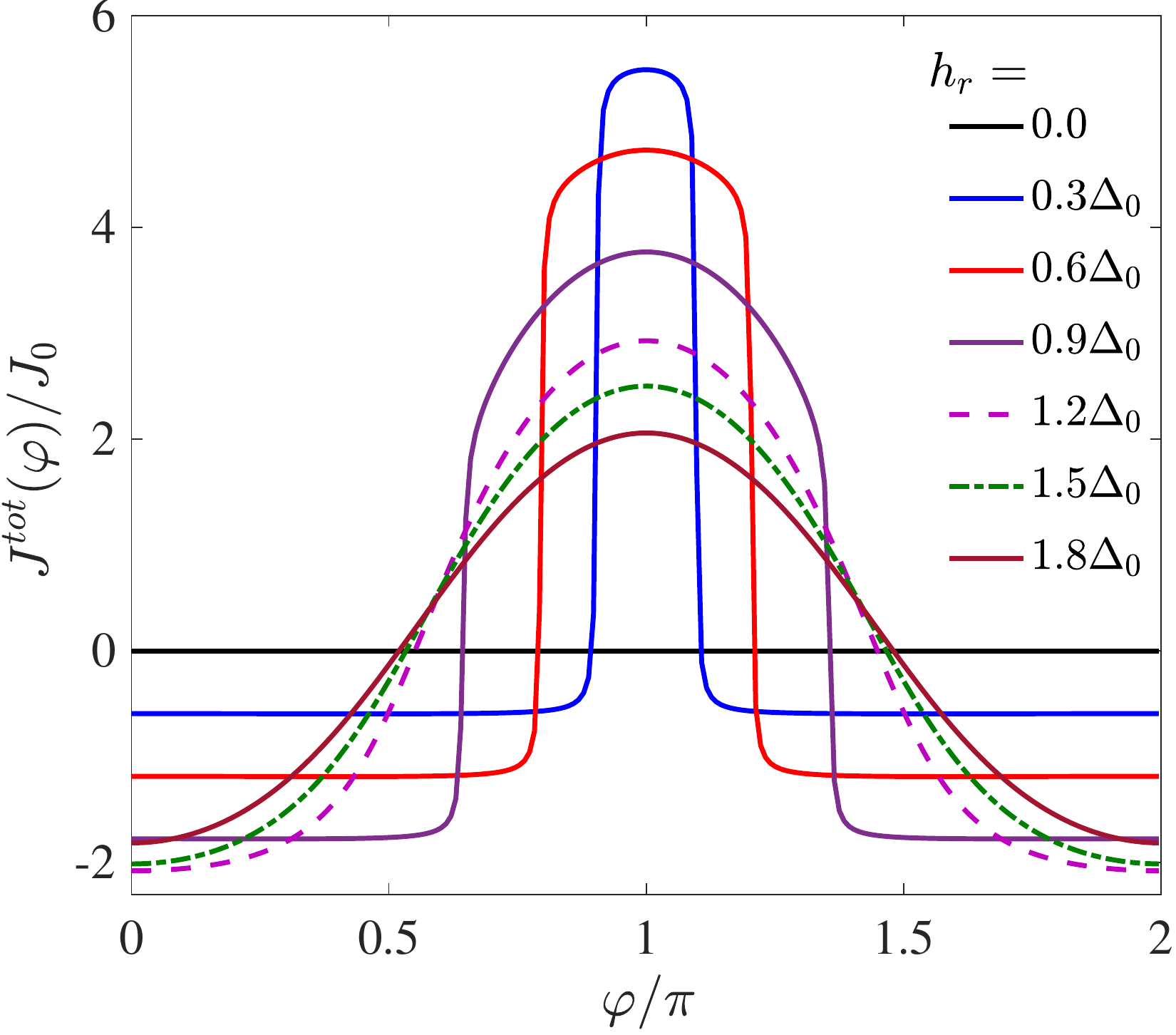}
\caption{ (Color online) Total supercurrent (spontaneous current) at
  the junction location $x=0$,
  displayed in Fig. \ref{fig1}, along the $y$ axis as a function of
  phase gradient perpendicular to the junction
  $\varphi=\theta_r-\theta_l$. The magnetization in the left segment
  is assumed zero $h_l=0$ while the magnetization in the right segment
  is directed along the $x$ axis normal to the interface and its strength varies from $h_r=0$
  to $1.8\Delta_0$.
 }\label{fig2}
\end{figure}

Let us return to the main structure we are interested in, namely,
a magnetized junction where superconductivity
is $s$-wave spin-singlet and not directional dependent. We first consider a simple case where
$\Delta_r=\Delta_l=\Delta$ and $h_l=0$. 
Figure \ref{fig2} exhibits the total spontaneous supercurrent
$J^{tot}$ at the junction
$x=0$ along the $y$ axis parallel to the interface as a function of
the superconducting phase difference between the left and right
segments $\varphi=\theta_r-\theta_l$. In our calculations, we have
assumed that the phase difference is controllable and uniform along the $y$ axis, and
neglected the influences of boundaries along the $y$ axis. The former
assumption can be understood by noting that the superconductivity is
induced to the surface states by an external electrode so that one can
control its macroscopic phase via the injection of a supercurrent. The
latter assumption is valid in a system where the junction is
wide enough compared to the superconducting coherence length so that
the boundaries in the $y$ direction are located in infinity\cite{zyuzin,bobkova,jap_alidoust,vrtx_ali,ali_so1,ali_so2}. As seen in
Fig. \ref{fig2}, $J^{tot}$ vanishes when the magnetization strength
is zero $h_r=0$. By increasing the magnetization strength, $J^{tot}$
is enhanced and eventually, further increase in $h_r$ suppresses $J^{tot}$.
The amplitude of the spontaneous
current reaches its maximum for all values of $h_r$ in phase
differences close to $\varphi=\pi$.
Also, the spontaneous current along the interface changes sign before and after $\varphi=\pi$. 
To understand the behaviors of
the spontaneous supercurrent $J^{tot}$ parallel to the interface at $x=0$, we
calculate the components of total current. To this end, the current
densities along opposite trajectories $A$ and $B$ parallel to the
interface, shown in Fig. \ref{fig1}, should be calculated so that the
total spontaneous current is given by $J^{tot}=J^A-J^B$. 
To simplify the current density phase relations and derive an
analytical expression, we set $\Delta_l=\Delta_r=\Delta$,
$h_l=0$, and $h_r\neq 0$ (however, in our numerical calculations, all
quantities are assumed nonzero that result in long expressions and
therefore we
avoid presenting them). The
supercurrent density along the $A$ trajectory is expressed by
\begin{equation}\label{jab}
 J=J_0 \sum_{n\in Z}\frac{2 {\cal Z} \Big\{{\cal S}_1 \sin
   (\varphi_0^r+\varphi)-\omega_n \sin \varphi+h_r
   \cos\varphi\Big\}}{{\cal D}},
\end{equation}
\begin{subequations}
\begin{eqnarray}
\nonumber &&{\cal D}=-2 \Big\{h_r
   \left({\cal Z}\sin
     \varphi-{\cal S}_1 \sin
\varphi_0^r\right)+{\cal S}_1\omega_n
   \cos \varphi_0^r \Big\}+\nonumber\\&& 2 {\cal S}_2{\cal Z}
   \cos \left(\varphi_0^r+\varphi\right)-2 \omega_n {\cal Z}
   \cos\varphi+ 2
   \omega_n \sqrt{\Delta ^2+\omega_n^2}+\nonumber\\&&\Delta ^2+h_r^2+\sqrt{\left(\Delta
   ^2+\omega_n^2\right)^2+h_r^4+2 h_r^2 \left(\omega_n^2-\Delta ^2\right)}+3
   \omega_n^2,~~~
\end{eqnarray}
\begin{eqnarray}
&& {\cal S}_1=\Big[2 \omega_n^2 \left(\Delta ^2+h_r^2\right)+\left(h_r^2-\Delta
   ^2\right)^2+\omega_n^4\Big]^{\frac{1}{4}},\\&&{\cal S}_2=\Big[\left(\Delta
   ^2+\omega_n^2\right)^2+h_r^4+2 h_r^2 \left(\omega_n^2-\Delta
   ^2\right) \Big]^{\frac{1}{4}},\\&&\varphi_0^r=\frac{1}{2} \arg \big[\Delta ^2-(h_r+i
  \omega_n)^2\big],\\&& {\cal Z}=\sqrt{\Delta ^2+\omega_n^2}+\omega_n. 
\end{eqnarray}
\end{subequations}
%\end{widetext}
\begin{figure}[t!]
\includegraphics[width=6.50cm,height=7.40cm]{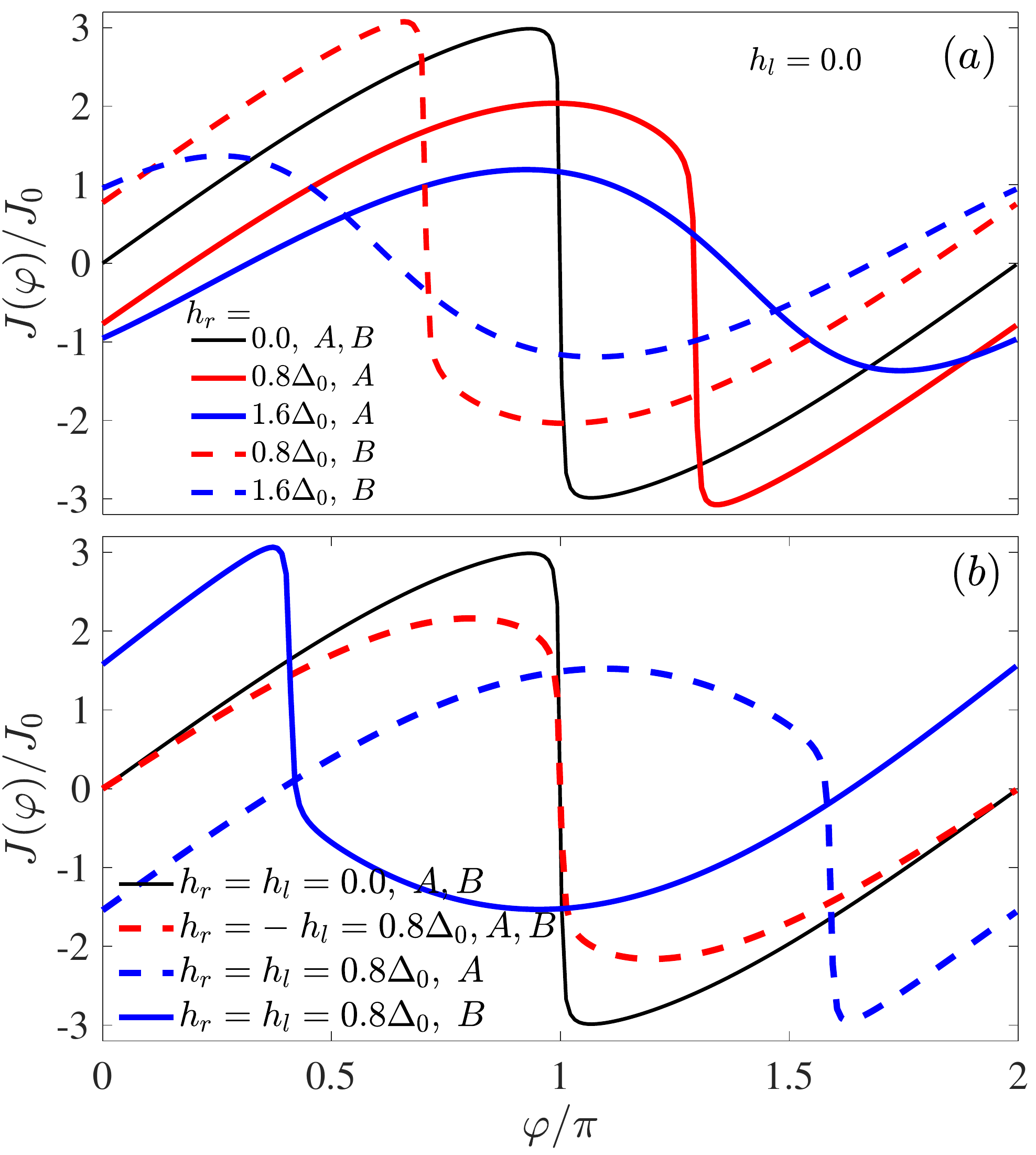}
\caption{ (Color online) Supercurrent densities along $A$ and $B$
  trajectories parallel to the interface shown in Fig. \ref{fig1} as a function of phase
  difference perpendicular to the junction interface $\varphi$. In
  panel (a) we set $h_l=0$ and vary $h_r$ while in panel (b) we
  consider situations where $h_r=\pm h_l$.}\label{fig3}
\end{figure}
\begin{figure*}[t!]
\includegraphics[width=14.70cm,height=4.10cm]{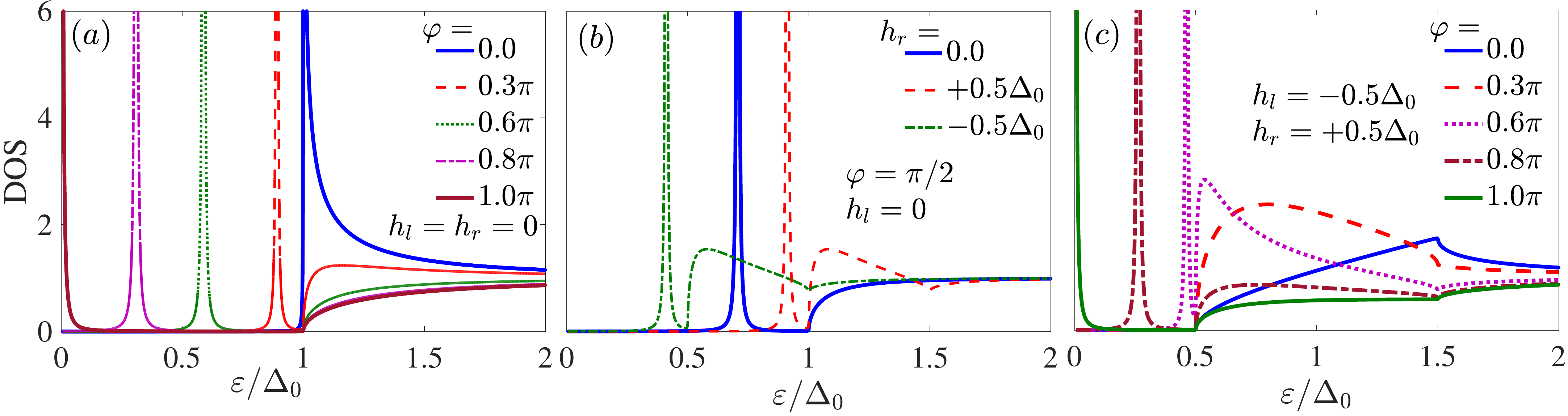}
\caption{ (Color online) The density of states DOS($\varepsilon$) as a function of the quasiparticles
  energy $\varepsilon$ at
  the interface of weak-link $x=0$. (a) We set $h_r=h_l=0$ and vary the phase
  difference $\varphi=0, 0.3\pi, 0.6\pi, 0.8\pi, \pi$. In panel (b),
  we examine the effect of magnetization direction on the Andreev
  subgap states by setting $h_l=0$, $\varphi=\pi/2$, and $h_r=0,\pm
  0.5\Delta_0$. (c) We consider opposite magnetization directions
  with identical intensities on
  the left and right sides of the weak-link $h_l=h_r=0.5\Delta_0$ and
  vary the phase difference similarly to panel (a).  
 }\label{fig4}
\end{figure*}
A phase relation similar to Eq. (\ref{jab}) can be obtained for the current density along the $B$ trajectory by
properly accounting for the magnetization direction. Figure \ref{fig3}(a) illustrates the spontaneous current densities
along trajectories $A$ and $B$ as a function of $\varphi$ for three
different values of $h_r=0, 0.8\Delta_0, 1.6\Delta_0$, and
$h_l=0$. Here we have defined $J_0=2e\pi TN(0)$. Note that the magnetization is oriented along the $x$ axis
perpendicular to the junction interface. As seen, the current
densities along the $A$ and $B$ trajectories are identical when
$h_r=0$. This is starkly oppose to the cases where $h_r\neq 0$. We see that the
current densities along the $A$ and $B$ trajectories are dissimilar
and therefore cause a finite spontaneous current along the interface
$J^{tot}=J^A-J^B\neq 0$. From Fig. \ref{fig3}(a) it is clear that
$J^{tot}$ is at a maximum at phases close to $\varphi=\pi$ due to the
nonsinusoidal
behaviour of the current density phase relations. The current
densities are nonzero at zero phase difference when $h_r\neq 0$; namely
the current density experiences a $\varphi_0$ phase shift in the presence of
magnetization. From Eq. (\ref{jab})
it is apparent that the current density is nonzero at zero phase
difference $\varphi=0$ when $h_r\neq 0$. It is worth mentioning that
the appearance of a $\varphi_0$ phase shift in the supercurrent has
theoretically been
discussed in various situations
\cite{Krive,zazan,mj,Heim,Feinberg,Goldobin,silaev1,silaev2,silaev3,Schrade,Shapiro,buz_so1,zyuzin,bobkova,vasenko2}
and observed in experiments \cite{roso,Szombati}. In structures where
the spin orbit meditated coupling is available, its interplay with a
properly oriented Zeeman(-like) field results in a supercurrent flow
perpendicular to the junction interfaces at zero phase
difference \cite{Schrade,Shapiro,buz_so1,zyuzin,bobkova,vasenko2}. Figure \ref{fig3}(b)
illustrates the current densities along the $A$ and $B$ trajectories
where the magnetizations in both sides of the weak-link are
nonzero. We see that when $h_r=-h_l$, i.e., the magnetizations are
equal in strength and oppositely
oriented perpendicular to the interface, the current
densities along the $A$ and $B$ trajectories are identical and hence
the total spontaneous current along the interface in the $y$ direction
vanishes similarly to the nonmagnetized case $h_r=h_l=0$. The
spontaneous current however reappears when $h_r=h_l$, namely, when the
magnetizations of both sides are oriented in the same direction and
perpendicular to the interface. From Fig. \ref{fig3}(b) it is clear
that the current densities as a function of $\varphi$ along the $A$
and $B$ trajectories are different if $h_r\neq 0$, and therefore it causes a finite
spontaneous current parallel to the interface. Comparing
Figs. \ref{fig3}(a) and \ref{fig3}(b), we conclude that a finite
$J^{tot}$ is also feasible even in a case where $h_r$ and $h_l$
have opposite orientations. The finite $J^{tot}$ in this case is
accessible when the magnetizations have different strengths,
i.e., $|{\bm h}_r|\neq |{\bm h}_l|$. Our numerical calculations (not shown) also confirm this fact.

One of the measurable physical quantities in the laboratory is the density
of states. The DOS can be detected by STM experiments or through
$I$-$V$ characteristic curves in a tunneling spectroscopy experiment where $dI/dV$ is proportional to the
DOS. The DOS in the quasiclassical approach is accessible through the
normal component of the total Green's function, i.e., Eq. (\ref{normalG}):
\begin{equation}\label{dos}
\text{DOS}(\varepsilon,{\bm r})=N(0)\underset{\delta\rightarrow 0}{\text{Re}}\Big\{g(i\omega_n\rightarrow \varepsilon+i\delta,{\bm r})\Big\},
\end{equation}
in which we have introduced an infinitesimal imaginary number $i\delta$
and, for convenience in our subsequent analyses, turn to the energy
representation by substituting $i\omega_n\rightarrow
\varepsilon+i\delta$. The imaginary part $i\delta$ helps to account
properly for the Green's function poles. In Fig. \ref{fig4} we plot the DOS as a function
of normalized quasiparticle energy $\varepsilon/\Delta_0$ at the
interface $x=0$. Figure
\ref{fig4}(a) illustrates the DOS where the phase difference $\varphi$ between
the two segments of the Josephson weak-link (see Fig. \ref{fig1}) is $0, 0.3\pi,
0.6\pi, 0.8\pi, \pi$. We also set $h_l=h_r=0$ which is equivalent to a
normal Josephson contact. At $\varphi=0$, the DOS shows the usual BCS
gap structure
with a singularity at $\varepsilon=\Delta_0$. When we set
$\varphi\neq 0$ a singularity appears at energies below the
superconducting gap $\varepsilon<\Delta_0$. This
singular point corresponds to an Andreev bound state due to the
resonance of particle-hole conversions at the interfaces of left and right
superconductors. The bound state moves to $\varepsilon=0$ when the
phase difference is maximum $\varphi=\pi$. In Fig. \ref{fig4}(b), we set $h_l=0$,
$\varphi=\pi/2$ and plot DOS for $h_r=0, \pm 0.5\Delta_0$. As seen,
the nonzero magnetization in the right segment of weak-link shifts the
Andreev subgap state to smaller or larger energies, depending on the
magnetization direction. If we set opposite magnetization directions
with identical strengths on opposite sides of the weak-link segments,
i.e., $h_l=-h_r$, the shift in the Andreev bound states induced by the
direction of magnetization disappears and the Andreev subgap state at
$\varphi=\pi$ reoccurs at $\varepsilon=0$.

To gain better insights, in what follows, we parametrize the Green's
function and derive an analytical expression for the Andreev bound states. To this
end, we make use of a so-called Riccati parametrization scheme
\cite{riccati} and define two propagators $\gamma$ and
$\tilde{\gamma}$ so that the Green's function is rewritten as follows:   
\begin{equation}
\check{g}=\frac{1}{1-\gamma\tilde{\gamma}}\left(\begin{array}{cc}
            1+\gamma\tilde{\gamma} &+2\gamma \\
            -2\tilde{\gamma} & -1-\tilde{\gamma}\gamma
          \end{array}\right).
\end{equation}
Substituting the parametrized Green's function into the
Eilenberger equation, Eq. (\ref{Eilenbrg}), two decoupled first-order
differential equations for $\gamma$ and $\tilde{\gamma}$ appear. After some
calculations, we find the following solutions to
$\gamma$ and $\tilde{\gamma}$ at the interface from the right side of the
weak-link $x\rightarrow 0^+$:
\small
\begin{subequations}
\begin{eqnarray}
\gamma_r=\frac{(\varepsilon+h_r)-\text{sgn}(\varepsilon+h_r)\sqrt{(\varepsilon+h_r)^2-\Delta_r^2}}{\Delta_r e^{-i\theta_r}},\\
\tilde{\gamma}_r=\frac{(\varepsilon+h_r)-\text{sgn}(\varepsilon+h_r)\sqrt{(\varepsilon+h_r)^2-\Delta_r^2}}{\Delta_r e^{+i\theta_r}}.
\end{eqnarray}
\end{subequations}
Similar solutions are derived to $\gamma$ and $\tilde{\gamma}$ at the
interface from $x\rightarrow 0^-$. The Andreev
bound states can be
determined through the singularities in the normal component of the
Green's function as discussed earlier [see Eq. (\ref{dos}) and
its associated results presented in Fig. \ref{fig4}]. Therefore,
the singularities
are solutions of $1-\gamma\tilde{\gamma}=0$ that result in:
\begin{eqnarray}
\cos\varphi-&&\frac{(h_l+\varepsilon)(h_r+\varepsilon)}{\Delta_l \Delta_r}+\nonumber\\&&\Bigg[\Big\{ 1-(\frac{h_l+\varepsilon}{\Delta_l})^2\Big\}\Big\{ 1-(\frac{h_r+\varepsilon}{\Delta_r})^2 \Big\}\Bigg]^{-\frac{1}{2}}=0,
\end{eqnarray}
where we invoked the left and
right indices $l,r$ for the quantities of the left and right segments
of the weak-link. By carrying out some calculations, we find the following relation to the Andreev bound states:
\begin{widetext}
\begin{equation}\label{abs}
\varepsilon_{\cal A}=\frac{h_r\Delta_l^2+h_l\Delta_r^2-{\cal
    H}_+\Delta_l\Delta_r\cos\varphi\pm\Delta_l\Delta_r\sin\varphi\sqrt{\Delta_l^2+\Delta_r^2-2
    \Delta_l\Delta_r\cos\varphi-{\cal H}_-^2}}{\Delta_l^2+\Delta_r^2-2
    \Delta_l\Delta_r\cos\varphi}
\end{equation}
\end{widetext}
in which we have defined ${\cal H}_\pm=h_l\pm h_r$ and $\varepsilon_{\cal A}$
determines the energy of the Andreev bound states. The supercurrent
flow passes through these subgap bound states. Hence, the associated
supercurrent phase relationship is proportional to the derivative
of the bound state energies with respect to the phase difference,
namely, $J\propto \sum_{\cal A}\frac{d\varepsilon_{\cal
    A}}{d\varphi}\tanh\beta\varepsilon_{\cal A}$ with $\beta=k_BT$.
Nonetheless, we do not calculate the supercurrent by this method and only focus our discussions on the analyses of the Andreev
bound states. To simplify the bound state expression Eq. (\ref{abs}), we first set
$\Delta_l=\Delta_r=\Delta$ and $h_r=h_l=0$ and consequently find
\begin{eqnarray}
\varepsilon_{\cal A}=\Delta\cos\varphi/2.
\end{eqnarray}
This relation shows that the bound state at $\varphi=0$ moves to the
edge of superconducting gap at $\varepsilon_{\cal A}=\Delta$ and to zero energy when $\varphi=\pi$
in line with previous works on the conventional Josephson short junctions \cite{Kulik,Bulaevskii,ko,golubov2}. These
results are consistent with our numerical calculations discussed in
Fig. \ref{fig4}($a$). We now set $h_l, h_r\neq 0$ and find the
following relation for the Andreev bound states:
\begin{eqnarray}
\varepsilon_{\cal A}=\frac{{\cal H}_+(1-\cos\varphi)\pm
  \sin\varphi\sqrt{2\Delta^2(1-\cos\varphi)-{\cal H}_-^2}}{2(1-\cos\varphi)}.
\end{eqnarray}
We see that the general aspects of the latter expression are in full
agreement with the numerical results presented in Figs. \ref{fig4}(b)
and \ref{fig4}(c). If we set $\varphi=\pi$, the bound state occurs at
$\varepsilon_{\cal A}=(h_l+h_r)/2$. It is evident that if the magnetization directions
in the left and right segments are oppose, $h_l=-h_r$, the bound state
takes place at $\varepsilon=0$ which is consistent with the DOS results
presented in Fig. \ref{fig4}(c). The
difference between the TI junction and a conventional one is the
presence of strong spin-orbit coupling (or equivalently the
spin-momentum locking), and therefore, the directional dependence
discussed above is a direct consequence of the spin-momentum locking
phenomenon. It is worth noting that not only can the DOS in an intrinsic spin-orbit
coupled magnetic superconducting hybrid be magnetization direction
dependent, but
also the charge and spin supercurrents are found to be sensitive to the direction of
magnetization \cite{dos_so,berg_so3,berg_so2,ali_so1,ali_so2}. The DOS
at maximum superconducting phase
difference $\varphi=\pi$ in a diffusive Josephson junction peaks at zero energy due to the appearance of
superconducting triplet correlations both in magnetic inhomogeneous \cite{dos_f}
and spin-orbit coupled systems\cite{dos_so}.

The spontaneous supercurrent explored here can be experimentally measurable
through multiterminal devices \cite{exp_hall}. Two transverse electrodes should
be attached to the lateral edges of the two-dimensional topological insulator weak-link at $x=0$ and
$y=\pm W/2$ where $W$ is the junction width and we assumed
$W\rightarrow\infty$ in our calculations (see Fig. \ref{fig1}). The transverse spontaneous
current parallel to the junction interface discussed above injects charge current into the lateral leads
and can induce a voltage drop between the lateral leads
that is detectable in experiment \cite{exp_hall}. By applying a
voltage difference between these lateral electrodes, the DOS and thus
the subgap bound states can be revealed in an $I$-$V$ measurement. When
these signatures are detected in an experiment, a rotatable in-plane
external magnetic field can confirm our findings. Our predictions are
valid regardless of the density/strength of nonmagnetic impurity and scattering
resources present at the surface channels. Also, to rotate the
magnetization in the setup proposed, an in-plane external magnetic field
suffices. Therefore, the impurity and Meissner obstacles pointed out in
the introduction to experimentally observe the spontaneous currents at the surfaces of
chiral superconductors are not relevant in the Josephson weak-link
considered in this paper.

\section{conclusions}\label{conclusions}

In conclusion, utilizing a recently generalized quasiclassical
approach to superconducting magnetized surface states of a
three-dimensional topological insulator (TI) \cite{zyuzin,bobkova}, we study
supercurrent flows at the magnetic interface of a TI. We consider a Josephson weak-link with two
uniformly in-plane magnetized segments, ${\bm h}_l$ and ${\bm h}_r$, where the magnetizations have nonzero
components perpendicular to the interface. Our results reveal that a
spontaneous supercurrent flows parallel to the interface at
the junction location provided that $|{\bm h}_l|\neq |{\bm h}_r|$ and reaches its maximum when the phase
difference $\varphi$ between the left and right segments is close
to $\pi$. We also study the Andreev bound states in such a weak-link
through the density of states both numerically and analytically. We
Riccati-parametrize the Green's function involved in our calculations
and derive analytical expressions to the Andreev subgap states. We
discuss the influences of the magnetization directions in the left and right sides
of the Josephson weak-link on the Andreev bound states. 

\acknowledgments
M.A. is thankful to I.V. Bobkova and A.M. Bobkov
for useful discussions.

\end{document}